\documentclass[]{article}
\usepackage{graphicx}
\usepackage[top=2.5cm, bottom=2.5cm, left=2cm, right=2cm]{geometry}
\usepackage{amssymb} 
\usepackage{amsmath}
\usepackage{epstopdf}
\usepackage[utf8]{inputenc}

\usepackage{subcaption}
\title{Optical manipulation of layer-valley coherence via strong exciton-photon coupling in microcavities} 
\date{\vspace{-5ex}}
\begin{document}
\author{Mandeep Khatoniar$^{1,2}$, Nicholas Yama$^{3}$, Areg Ghazaryan$^{4}$, Sriram Guddala$^2$, Pouyan Ghaemi$^{1,2}$,\\Kausik Majumdar$^5$, Vinod Menon$^{1,2}$}
\maketitle

\begin{center}
$^1$ Department of Physics, The Graduate Center, City University of New York, New York 10016, United States\\
$^2$ Department of Physics, City College of New York, City University of New York, New York 10031, United States\\
$^3$ University of Hawai'i at Manoa, 2500 Campus Rd, Honolulu, HI 96822\\
$^4$ IST Austria (Institute of Science and Technology Austria), Am Campus 1, 3400 Klosterneuburg, Austria\\

$^5$ Department of Electrical and Computer Engineering, Indian Institute of Science, CV Raman Rd, Bengaluru, Karnataka 560012, India\\ Corresponding Author: vmenon@ccny.cuny.edu.

\end{center}
\begin{abstract}
Coherent control and manipulation of quantum degrees of freedom such as spins forms the basis of emerging quantum technologies. In this context, the robust valley degree of freedom and the associated valley pseudospin found in two-dimensional transition metal dichalcogenides is a highly attractive platform. Valley polarization and coherent superposition of valley states have been observed in these systems even up to room temperature. Control of valley coherence is an important building block for the implementation of valley qubit. Large magnetic fields or high-power lasers have been used in the past to demonstrate the control (initialization and rotation) of the valley coherent states. Here we demonstrate control of layer-valley coherence via strong coupling of valley excitons in bilayer WS$_2$ to microcavity photons by exploiting the pseudomagnetic field arising in optical cavities owing to the TE-TM splitting. The use of photonic structures to generate pseudomagnetic fields which can be used to manipulate exciton-polaritons presents an attractive approach to control optical responses without the need for large magnets or high intensity optical pump powers.         
\end{abstract}

The need for efficient and fast transfer of information has driven a large body of research in the past decades that lead to finding alternatives to using moving electrons. Utilizing the spin of the electrons as means of information storage or Spintronics emerged as a result of such research but difficulty in transfer and control of electronic spin and the cryogenic temperatures or complicated sample preparations for maintaining long spin coherence times leave much room for development in this field  [1]. Transition Metal Dichalcogenides (TMDCs)are a class of layered van der Waals (vdW) materials which in addition to various electronic and optical properties like a direct band gap at the Brillouin zone edges (K and K’ valleys) and pronounced spin orbit coupling, have come into limelight for their valley selective optical transition rules arising from the breaking of inversion symmetry and their coupling of the layer, valley and spin degrees of freedom  [2].\\
These valley selective transitions have been exploited to demonstrated valley polarization under circularly polarized pump which excited carriers in either the K or K’ valleys  [3] and valley coherence under linear polarized excitation where both valleys are excited [4]. Valley coherence has received much interest in the context of using the valley pseudospin as a qubit [5] and realizing entanglement between the two valleys when excited with single photons [6]. Till date the manipulation of valley coherence has been achieved either using magnetic fields [7] or using intense optical fields [8,9].\\
The TMDCs have also become a very attractive platform for studying strong light-matter coupling and exciton-polariton formation [10]. Indeed their valley degree of freedom was exploited in a number of recent works on strong coupling in these low-dimensional systems  [11–14]. More recently the enhancement in valley coherence under strong coupling and the control of the valley exciton-polariton propagation was also demonstrated [15–17].While monolayer TMDCs did show valley coherence, these observations were mostly at low temperature to suppress the undesirable intervalley scattering. In addition to the valley degree of freedom, the 2D TMDCs also offer the layer and spin degrees of freedom [18,19]. The interplay between these degrees of freedom results in robust layer-spin-valley coherence which was demonstrated in bilayer TMDCs [20]. \\
Here we demonstrate the control of layer-valley coherence in bilayer WS$_2$ via strong coupling of excitons to cavity photons by exploiting the pseudomagnetic field that arises in optical cavities owing to the TE-TM splitting of the cavity photon mode [21,22].

\begin{figure}[h!]
\begin{subfigure} [h!]{\textwidth}
\centering
\includegraphics[width=0.8\textwidth]{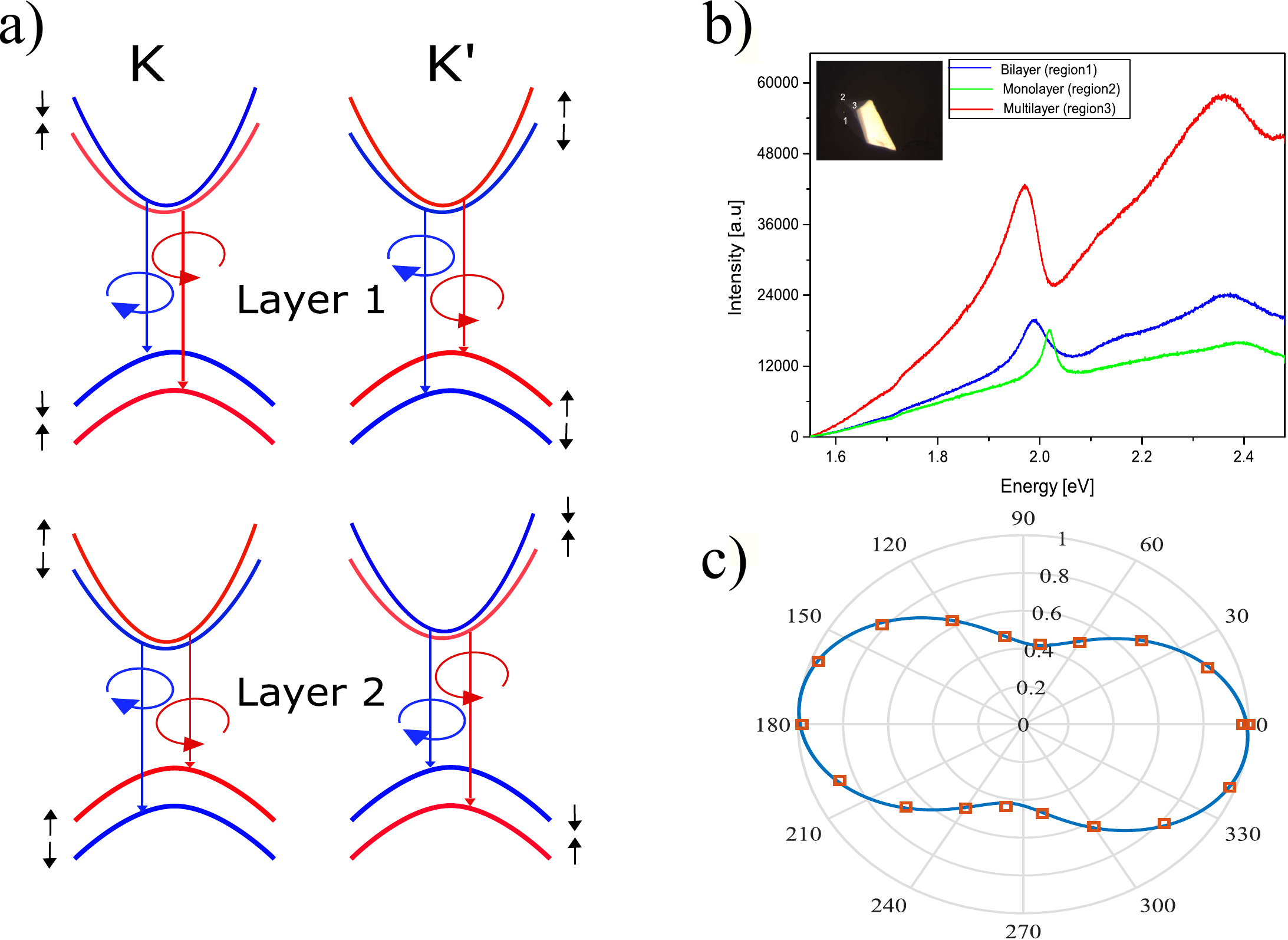}
\end{subfigure}
\caption{ Fig 1a shows the coherence scheme in bilayer WS$_2$. Fig 1b shows the white light spectrum comparing different TMDC layers with inset showing optical image of the bilayer sample used for the experiment. Fig 1c. shows the response of the bilayer photoluminescence (PL) under rotation of the analyzer with laser polarization horizontal to the slit direction of the spectrometer.}
\end{figure}
Fig. 1a is a cartoon showing the bands at K and K’ valleys for the tungsten (W) based bilayer TMDC system. Here the helicity dependent optical selection rules get flipped between the two layers as shown by the red and blue transitions in layers 1 and 2 due to their 2H stacking. Although restoration of center of inversion indicates the loss of valley dichroism, the bilayer system retains its spin dichroism as the spin up and spin-down states are still decoupled owing to the interlayer hopping that conserves spin [25]. This has been realized experimentally in [20] where bilayer WS$_2$ retain robust degrees of valley polarization and valley coherence even at room temperature. Bilayer WS$_2$ was obtained via mechanical exfoliation and were identified using white light reflection as shown in Fig. 1b. After transferring on a quartz substrate linear polarization resolved photoluminescence(PL) measurements were done with a near-resonant excitation of 620 nm using tunable pulsed laser. All measurements were done at room temperature unless specified otherwise(For details of fabrication and experimental methods - See \emph{Methods}). The degree of linear polarization (DOP) is defined by $\rho =\frac{I_{max}-I_{min}} {I_{max}+I_{min}}$, where I$_{max}$ is the maximum intensity and I$_{min}$ is the minimum intensity obtained as after the PL passes through an analyzer. The value of $\rho$ was obtained by fitting the obtained analyzer angle $\Theta$ and normalized intensity I to $I=cos^2{(\Theta-\phi})+\frac{1-\rho}{1+\rho}sin^2({\Theta-\phi})$ with $\rho$ and $\phi$ (rotation of plane of polarization) as the fitting parameters.For the bilayer sample on quartz cover slip a value of $\rho= 0.42$ was obtained which agrees with previously reported value[20]. Non-resonant excitation at 532nm shows  a lower DOP of $\rho=0.29$ (See \emph{Supplementary} Fig S1) due to increased chances of  scattering and electron hole exchange [24]. Coherence measurements were also performed on monolayer WS$_2$ on quartz which yielded a $\rho=0.14$ which is low compared to the bilayer (see \emph{Supplementary} Fig S2). Thus a high degree of coherence retention from the two momentum separated valleys is observed as shown in Fig. 1c, thereby indicating that bilayer WS$_2$ is robust against aforementioned depolarization mechanisms present in these systems which serves as the main motivation for the use of bilayer $WS_2$ in the polariton system over monolayers, where one has to go to low temperatures to observe an appreciable degree of valley coherence[15].
\begin{figure}[h!]
\begin{subfigure} [h!]{\textwidth}
\centering
\includegraphics[width=0.8\textwidth]{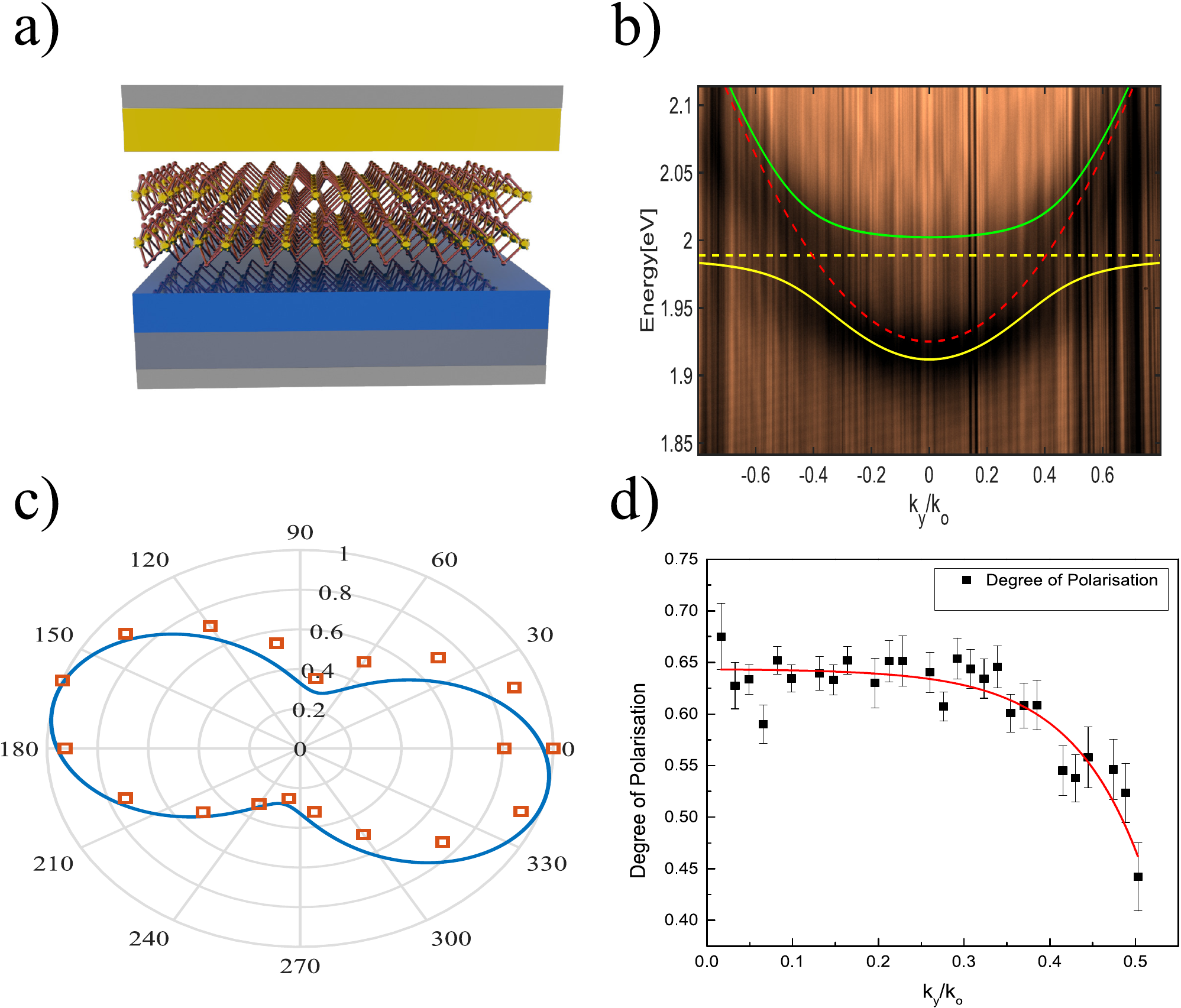}

\end{subfigure}
\caption{Fig 2a. shows the cavity schematics used in the experiments. Fig 2b. shows the white light dispersion of the microcavity polaritons.The green and yellow solid lines correspond to the upper and lower polariton branches obtained via coupled oscillator fit. The dashed yellow line is the bilayer WS$_2$ exciton resonance and the red dashed line is the cavity at dispersion. Fig 2c. shows the response of the polariton photoluminescence (PL) under rotation of the analyzer with laser polarization horizontal to the slit direction of the spectrometer. Fig 2d. shows the dependence of the DOP with increasing k$_\parallel$}.
\end{figure} 
\paragraph{}The bilayer was then encapsulated in an all Ag metal cavity with Al$_2$O$_3$ and PMMA as a spacer layer as shown in Fig. 2a (See \emph{Methods}). Fig. 2b shows the white light reflection of the encapsulated bilayer. The upper polariton (UP) and lower polariton(LP) branch are fitted to a coupled oscillator model (indicated by the green and yellow solid lines respectively) to yield a Rabi splitting of $\emph{g}\sim 40$ meV with a cavity detuning of $\sim61.5$ meV with respect to the exciton line . polarization resolved PL fourier space image of the bilayer was taken with a 620nm pulsed laser aimed a resonant excitation of the UP branch at k$_\parallel=0$ photon in-plane momentum(denoted in the figures as k$_y/$k$_0$). Here $k_\parallel = \frac{2\pi}{\lambda}sin\alpha$ with $\lambda$ as the wavelength and $\alpha$ as the angle made by the incident k (momentum) vector with the normal. Cross sections on the PL image were binned at different intervals along the slit direction (k$_y$) and the intensities were plotted against the analyzer rotation angles. Fig. 2c shows the polar plots with their respective fits at k$_\parallel= 0$ . A DOP of $\rho=0.65$ is obtained for the polariton emission at low in-plane momentum which is slightly higher than the bare bilayer emission. This attributed to the low lifetime of the polariton state at such low k$_\parallel$. The polariton lifetime at a specific k$_\parallel$ is given by $\tau_k=\frac{\tau_{photon}}{C_k}$, where $\tau_{photon}$ is the lifetime of the photon inside the cavity as determined by it's quality factor and $C_k$ is the photon fraction of the polariton at a given in-plane momentum. At small k$_\parallel$ the photon fraction of the cavity is the highest and hence the lifetime of the polariton is reduced . This process competes with the depolarization processes thereby allowing the polariton to retain a higher degree of coherence compared to that of the uncoupled exciton.. A gradual reduction in the DOP obtained after fitting the data is observe as shown in Fig. 2d. Error bars indicate the error in the Malu's law fittings. This is attributed to the fact that at high in-plane momentum the lifetime of the polariton increases as the lower polariton branch acquires a higher excitonic fraction. The calculated $\tau$ for k$_\parallel= 0$ and k$_\parallel=0.5k_0$ are 53 fs and 98 fs respectively for a cavity of quality factor of 70. Also the TE-TM splitting becomes significant with increase in k$_\parallel$ which provide an additional spin relaxation pathway for the polaritons. 
\begin{figure}[h!]
\begin{subfigure} [h!]{\textwidth}
\centering
\includegraphics[width=0.8\textwidth]{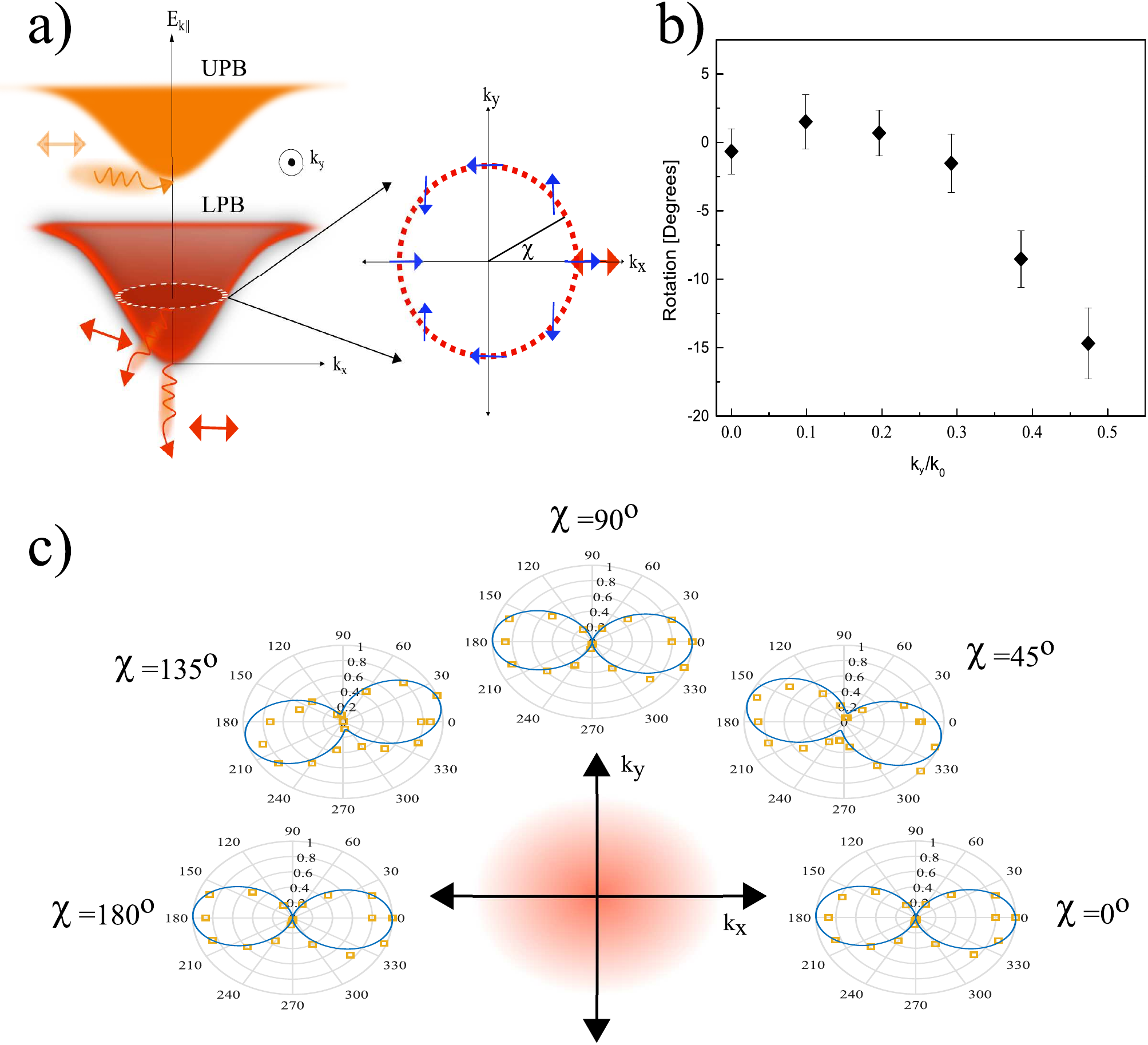}
\end{subfigure}
\caption{Fig 3a. shows the schematic of excitation with a horizontally polarized laser at k$_\parallel=0$ for the UPB and the rotation of plane of polarization inside the microcavity at a momentum of k$_\parallel=0.5k_o$ marked by the white dashed circle. The double headed arrow indicate the polarization of the input (orange) and the output PL at different in-plane momentum (red). The pseudo-magnetic field associated with that in-plane momentum is shown with blue arrows indicating the direction of the pseudo-magnetic field and the red double headed arrow showing the initialization of the polariton pseudospin . Fig 3b. shows the radial dependence of the rotation of polarization $\phi$ as a function of k$_\parallel$ at a fixed azimuthal angle $\chi{_k} = 45^o$. Error bars indicate the errors due to Malu's law fits . Fig 3c. Shows the rotation of the plane of polarization $\phi$ as a function of azimuthal angle of the wave vector $\chi{_k}$ with a constant k$_\parallel= 0.5k_0$}
\end{figure} 
\paragraph{} Next we demonstrate to manipulation of the plane of polarization with the aid of the microcavity. Fig. 3a shows the schematic of process by which one can realise this rotation in a microcavity taking advantage of the LT mode splitting which occurs due to difference in effective path length traversed by the L(TE) and the T (TM) modes inside a microcavity [21] .The presence of an LT splitting gives rise to an effective magnetic field(pseudomagnetic field) that goes as B $\sim (k_x\pm ik_y)^2e^{2i\chi_k}$, where $k_x$ and and $k_y$ are the x and y component of the in-plane wavevector k . The presence of this magnetic field causes the precession of the polariton psuedospin when initialised with a non zero angle with respect to the magnetic field. The z component of the psuedospin acquired due to the rotation decays as it moves along the poincare sphere till it is emitted as the polariton decays. And the projection of the the psuedospin state on the k$_x\emph{-}$k$_y$ plane just before it decays determines it's plane of polarization[22-23 \textit{and the references therein}]. To this purpose a polarization resolved full Fourier image of the emission is obtained by removing the slit and letting the Fourier image fall on the CCD directly bypassing the grating spectrometer(See \emph{Methods}). A fixed horizontal input polarization is maintained throughout the experiment . Fig. 3b shows the rotation of the plane of polarization as a function of k$_\parallel$ .Fig. 3c shows the polar plots of emission intensity as a function of analyzer angle at a fixed value of k$_\parallel \sim 0.5$ but with varying azimuthal angle $\chi$. The rotations are limited by the lifetime of the polariton species and the strength of the effective magnetic field. The Z component acquired as a result of the precession decays as a function of $\frac{2k_xk_y e^{\frac{-t}{\tau}}}{(k_x-ik_y)^2}$. As the elastic circle increases with increasing k$_\parallel$ the strength of the pseudo-magnetic field increases and thus the z component of of the psuedospin decays faster. Hence we have an upper limit to the range of rotations available to us . One could change the dielectric environment in the cavity spacer or increase the polariton lifetime via enhanced quality factor to increase the extent of rotation available to us. We perform a density matrix formalism to simulate the results obtained which is shown below.

The polarization dynamics of lower polaritonic branch can be described using density matrix formalism [26]. We define density matrix as $\rho_k\left(t\right)= \frac{1}{2}N_k\left(t\right)+S_k\left(t\right)\cdot\sigma$ where $N_k\left(t\right)$ is the total population of the polaritons at the lower branch. $S_k\left(t\right)$ and $\sigma$ are the pseudospin vector and the Pauli matrix acting on the states excited by the right and left circularly polarized light . Disregarding the hopping between the layers, each layer acts as an independent monolayer of TMD, with valley indices being reversed for two layers. Therefore, in contrast to monolayer, there is no valley dichroism for bilayer system and specific absorption of circularly polarized light solely corresponds to the spin dichroism close to the band edge[20,25] . Due to the large TE-TM splitting of the metallic cavity and extremely short lifetime of the cavity photon, we assume that the photon population and pseudospin dynamics of the polaritons are determined by their photonic component. Therefore, the dynamics of the population and the pseudospin are governed by 
\begin{equation}
\frac{dN_k}{dt}=-\frac{1}{\tau_k}N_k+P_k
\end{equation}
\begin{equation}
\frac{dS_k}{dt}=-\frac{1}{\tau_{sk}} S_k+\left[\Omega_k^i\times S_k\right]+\frac{P_k}{2}
\end{equation} 
 Where $\tau_k$ is the polariton lifetime as defined previously in the text.$\Omega_k^i = \left(\Omega_k^x,\Omega_k^y,0\right)= \Omega_k \left( cos\left(2\chi_k \right),sin\left(2\chi_k \right),0\right)$ is the pseudo-magnetic field due to the TE-TM splitting of the cavity modes $\Delta_{LT}\left(k\right)\left(i.e. \Omega_k = \frac{{\mid C_k\mid^2 \Delta_LT\left(k\right)}}{\hbar}\right)$,$C_k$ is the polariton photon fraction and $\tau_{sk}$ is the pseudospin lifetime resulting from the decoherence mechanisms other than the cavity pseudomagnetic field.$\chi_k$ is the azimuthal angle of the momentum $k_\parallel$ and $P_k$ describes the polariton flux initialized by the pump. In the current setup pump is linearly polarized along x axis, $i.e. P_k=(P_k,0,0)$. In steady state,we get
 \begin{equation}
 N_k=P_k\tau_k
\end{equation} 
 \begin{equation}
 S_k^z=\frac{-P_k\tau_{sk}^2\Omega_k\sin{2\chi_{k}}}{2(1+\Omega_k^2\tau_k^2)}
 \end{equation}
 \begin{equation}
 S_{k}^x=\frac{P_k\tau_{sk}(1+\Omega_k^2\tau_{sk}^2\cos^2(2\chi_{k})}{2(1+\Omega_k^2\tau_{sk}^2)}
 \end{equation}
 \begin{equation}
 S_{k}^y=\frac{P_k\tau_{sk}^3\Omega_k^2\sin\left(4\chi_{k}\right)}{4(1+\Omega_k^2\tau_{sk}^2)}
 \end{equation}
 The plane of polarization angle is given by
 \begin{equation}
 \phi=\frac{1}{2}\arctan{\frac{S_{k}^y}{S_{k}^x}}=\frac{1}{2}\arctan{\frac{\Omega_k^2\tau_{sk}^2\sin\left(4\chi_{k}\right)}{2(1+\Omega_k^2\tau_{sk}^2\cos^2(2\chi_{k})}}
 \end{equation}
 \begin{figure}[h!]
\begin{subfigure} [h!]{\textwidth}
\centering
\includegraphics[width=0.8\textwidth]{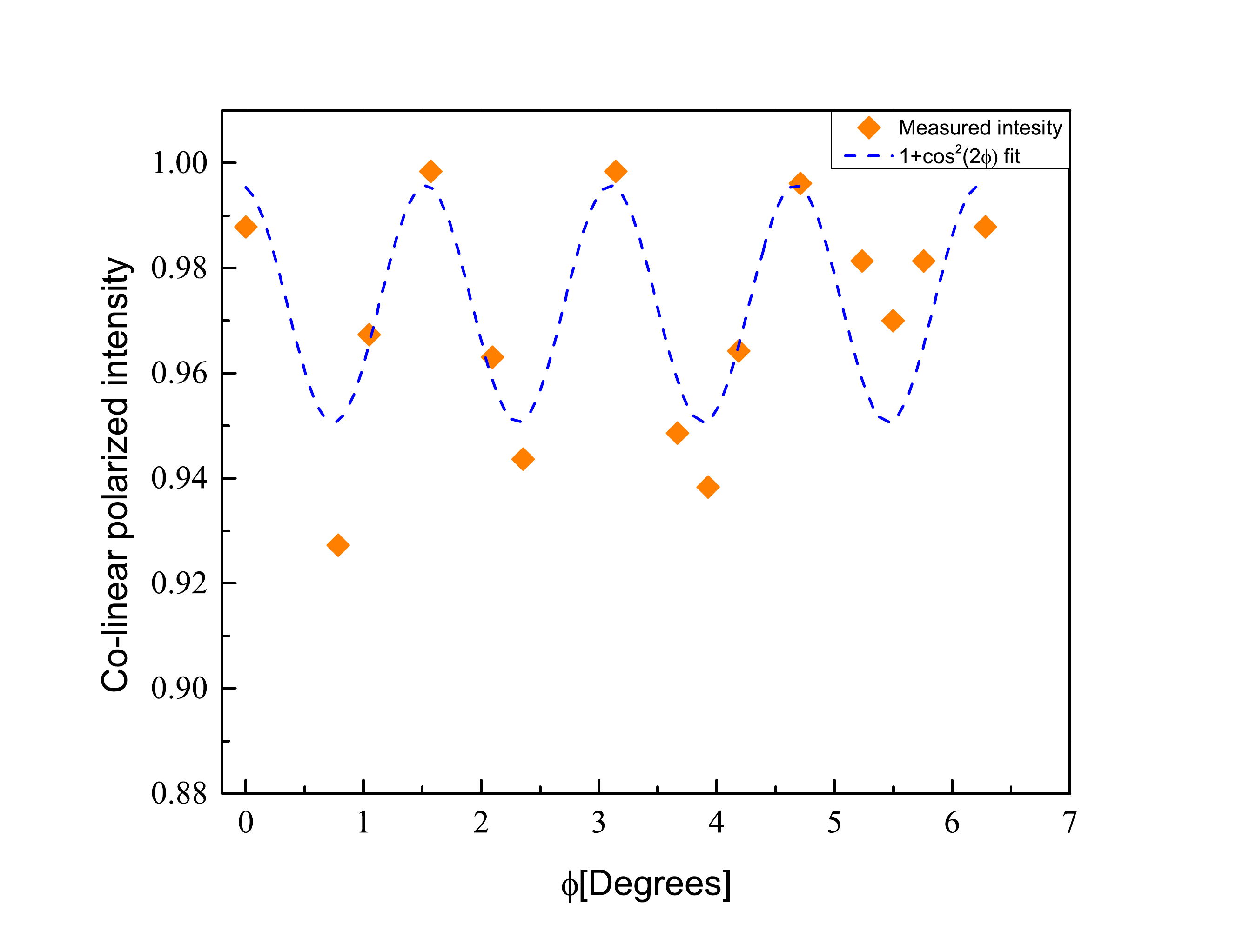}
\end{subfigure}
\caption{ Fig 4 shows the the co-linear intensity as a function of the azimuthal angle $\chi$ of the momentum space circle. The orange squares are measured values where the laser polarization and the analyzer are co-linear. The blue line shows a $1+cos^2(2\chi)$ dependence as obtained from equation (5).  }
\end{figure}
\\Fig. 4 shows the dependence of the population co-polarized with that of the laser, with a pseudospin corresponding to $S_{k}^x$, and angle of rotation of the polarization as a function of the azimuthal angle $\chi_{k}$. In order to obtain values, close to the experiment, we had to set $\tau_{sk}={0.9\tau}_k$, which indicates that the dominant mechanism of decoherence is the rotation of the pseudospin due to the pseudomagnetic field of TE-TM splitting of the optical cavity. The values of the angle of rotation of polarization are close to the experimentally obtained result, while all three quantities shown in equations (4),(5) and (6) (See \textit{Supplementary} Fig S5 for plots) perform four full cycles during the $2\pi$ change of $\chi_{k}$. The rotation of plane of polarization $\phi$ as seen from equation (7) and Fig S5 b, show that under the current system conditions the maximum rotation available to us is approximately 15 degrees. Similar behavior was also observed for AlGaAs quantum well systems at cryogenic temperatures using time resolved dynamics of the exciton-polaritons[23].

In summary, we demonstrate room temperature strong coupling of bilayer WS$_2$ and enhanced retention of layer-valley coherence of the polaritons. We exploit the pseudomagnetic field arising from the TE-TM splitting in an optical microcavity to manipulate the polariton layer-valley coherence. The possibility to control the polariton pseudospin by utilizing the pseudomagnetic field generated by the optical anisotropy in cavities presents a hitherto unexplored approach for control of layer-valley coherence in 2D materials via dispersion engineered photonic systems. This presents a promising route to realize active control of valley coherence such as for qubit applications where on-demand initialization, rotation and readout of the valley pseudospin is desirable. Optical cavities and photonic nanostructures with reconfigurable dispersion will be of even greater advantage in this context to realize on demand TE-TM splitting of desired value.
\section*{Methods}
\emph{Sample Preparation}:Films of bilayer WS$_2$ were produced via mechanical exfoliation of a single bulk crystal (acquired from HQ Graphene) and identified using photoluminescence and reflectance spectroscopy. The bilayer samples were then transferred, via Polydimethylsiloxane (PDMS) stamping, onto a 100 nm silver deposited substrate coated with a thin aluminum oxide film of 65 nm deposited via Atomic Layer Deposition(ALD) serving as the bottom mirror and dielectric spacer respectively. The samples were then annealed in argon, spin coated with poly(methyl methacrylate) (PMMA)which serves as a top spacer. A top layer of silver was deposited using electron beam evaporation. Strong coupling could then be verified using wavevector resolved photoluminescence and reflectance spectroscopy.\\
\emph{Experimental Setup}:Optical characterization was performed using a Princeton Monochromator and PIXIS CCD camera, with excitation from a tunable (500fs,80MHz) pulsed laser (TOPTICA Photonics), mounted onto an Olympus microscope in reflection geometry. A 50 micron pinhole was placed at the Fourier image plane prior to the sample to block incident light with large in-plane momenta. The output beam was sent through a variable polarizer and lens situated at the Fourier plane resolving the PL into momentum space. In wavelength-ky resolved spectra (for k-dependence of DOP) a slit along the ky direction and diffraction grating were used prior to imaging the spectra.In contrast, for full momentum-space images (for observation rotations) the slit and diffraction grating were removed.\\
\emph{Methods of Analysis}:The resulting images were then be processed in MATLAB. In the case of wavelength resolved images, the intensities of small intervals of k-values at the lower polariton emission were summed and averaged. Similarly, for the full k-space images, small squares centered at selected probing points in k-space were summed and averaged. In either case, the resulting data points could be fitted to a modified Malus’s law equation to determine both the DOP and angle of rotation .

\section*{Acknowledgements}
Authors acknowledge insightful discussions with Wang Yao. MK, NY, VM acknowledges NSF grants NSF DMR-1709996, and NSF OMA 1936276 for support. SG acknowledge the Army Research Office Multidisciplinary University Research Initiative program (W911NF-17-1-0312) for support. Authors acknowledge Nanofabrication facility at the Advanced Science Research Center where the fabrication of the devices were performed. Authors declare no competing interest. 
\section*{Bibliography}
1. 		S. D. Bader and S. S. P. Parkin, "Spintronics," Annu. Rev. Condens. Matter Phys. 1, 71–88 (2010).\\
2. 		J. R. Schaibley, H. Yu, G. Clark, P. Rivera, J. S. Ross, K. L. Seyler, W. Yao, and X. Xu, "Valleytronics in 2D materials," Nat. Rev. Mater. 1, 16055 (2016).\\
3. 		H. Zeng, J. Dai, W. Yao, D. Xiao, and X. Cui, "Valley polarization in MoS 2 monolayers by optical pumping," Nat. Nanotechnol. 7, 490–493 (2012).\\
4. 		A. M. Jones, H. Yu, N. J. Ghimire, S. Wu, G. Aivazian, J. S. Ross, B. Zhao, J. Yan, D. G. Mandrus, D. Xiao, W. Yao, and X. Xu, "Optical generation of excitonic valley coherence in monolayer WSe2.," Nat. Nanotechnol. 8, 634–8 (2013).\\
5. 		Y. Wu, Q. Tong, G.-B. Liu, H. Yu, and W. Yao, "Spin-valley qubit in nanostructures of monolayer semiconductors: Optical control and hyperfine interaction," Phys. Rev. B 93, 045313 (2016).\\
6. 		M. Tokman, Y. Wang, and A. Belyanin, "Valley entanglement of excitons in monolayers of transition-metal dichalcogenides," Phys. Rev. B - Condens. Matter Mater. Phys. 92, (2015).\\
7. 		G. Aivazian, Z. Gong, A. M. Jones, R.-L. Chu, J. Yan, D. G. Mandrus, C. Zhang, D. Cobden, W. Yao, and X. Xu, "Magnetic control of valley pseudospin in monolayer WSe2," Nat. Phys. 11, 148–152 (2015).\\
8. 		E. J. Sie, J. W. McIver, Y.-H. Lee, L. Fu, J. Kong, and N. Gedik, "Valley-selective optical Stark effect in monolayer WS2," Nat. Mater. 14, 290–294 (2015).\\
9. 		Z. Ye, D. Sun, and T. F. Heinz, "Optical manipulation of valley pseudospin," Nat. Phys. 13, 26–29 (2017).\\
10. 	X. Liu, T. Galfsky, Z. Sun, F. Xia, and E. Lin, "Strong light-matter coupling in two-dimensional atomic crystals," Nat. Photonics 9, 30–34 (2015).\\
11. 	Z. Sun, J. Gu, A. Ghazaryan, Z. Shotan, C. R. Considine, M. Dollar, B. Chakraborty, X. Liu, P. Ghaemi, S. Kéna-Cohen, and V. M. Menon, "Optical control of room-temperature valley polaritons," Nat Phot. 11, 491–496 (2017).\\
12. 	Y.-J. Chen, J. D. Cain, T. K. Stanev, V. P. Dravid, and N. P. Stern, "Valley-polarized exciton–polaritons in a monolayer semiconductor," Nat. Photonics 11, 431–435 (2017).\\
13. 	DufferwielS., L. P., S. D., T. A. P., WithersF., SchwarzS., MalpuechG., S. M., N. S., S. S., K. N., and T. I., "Valley-addressable polaritons in atomically thin semiconductors," Nat Phot. 11, 497–501 (2017).\\
14. 	N. Lundt, P. Nagler, A. Nalitov, S. Klembt, M. Wurdack, S. Stoll, T. H. Harder, S. Betzold, V. Baumann, A. V Kavokin, C. Schüller, T. Korn, S. Höfling, and C. Schneider, "Valley polarized relaxation and upconversion luminescence from Tamm-plasmon trion–polaritons with a MoSe2 monolayer," 2D Mater. 4, 25096 (2017).\\
15. 	S. Dufferwiel, T. P. Lyons, D. D. Solnyshkov, A. A. P. Trichet, F. Withers, G. Malpuech, J. M. Smith, K. S. Novoselov, M. S. Skolnick, D. N. Krizhanovskii, and A. I. Tartakovskii, "Valley coherent exciton-polaritons in a monolayer semiconductor," (2018).\\
16. 	L. Qiu, C. Chakraborty, S. Dhara, and A. N. Vamivakas, "Room-temperature valley coherence in a polaritonic system," Nat. Commun. 10, 1513 (2019).\\
17. 	N. Lundt, Ł. Dusanowski, E. Sedov, P. Stepanov, M. M. Glazov, S. Klembt, M. Klaas, J. Beierlein, Y. Qin, S. Tongay, M. Richard, A. V. Kavokin, S. Höfling, and C. Schneider, "Optical valley Hall effect for highly valley-coherent exciton-polaritons in an atomically thin semiconductor," Nat. Nanotechnol. 14, 770–775 (2019).\\
18. 	H. Yu, X. Cui, X. Xu, and W. Yao, "Valley excitons in two-dimensional semiconductors," Natl. Sci. Rev. 2, 57–70 (2015).
19. 	S. Das, G. Gupta, and K. Majumdar, "Layer degree of freedom for excitons in transition metal dichalcogenides," Phys. Rev. B 99, 165411 (2019).\\
20. 	B. Zhu, H. Zeng, J. Dai, Z. Gong, and X. Cui, "Anomalously robust valley polarization and valley coherence in bilayer WS2.," Proc. Natl. Acad. Sci. U. S. A. 111, 11606–11 (2014).\\
21. 	I. A. Shelykh, A. V Kavokin, Y. G. Rubo, T. C. H. Liew, and G. Malpuech, "Polariton polarization-sensitive phenomena in planar semiconductor microcavities," Semicond. Sci. Technol. 25, 013001 (2010).\\
22. 	O. Bleu, D. Solnyshkov, and G. Malpuech, "Optical Valley Hall Effect based on Transitional Metal Dichalcogenide cavity polaritons," (2016).\\
23.		W. Langbein, I. Shelykh, D. Solnyshkov, G. Malpuech, Yu. Rubo, and A. Kavokin, "Polarization beats in ballistic propagation of exciton-polaritons in microcavities," Phys. Rev. B 75, 075323 (2007).\\
24.		Hao, K., Moody, G., Wu, F. et al. "Direct measurement of exciton valley coherence in monolayer WSe2". Nature Phys 12, 677–682 (2016).\\
25.		Gong, Z., Liu, GB., Yu, H. et al. "Magnetoelectric effects and valley-controlled spin quantum gates in transition metal dichalcogenide bilayers". Nat Commun 4, 2053 (2013).\\
26.		Shelykh, I.A., Kavokin, A.V. and Malpuech, G., "Spin dynamics of exciton polaritons in microcavities". phys. stat. sol. (b), 242: 2271-2289 (2005)

\end{document}